# The surface-tension-driven Bénard conventions and unique sub-grain cellular microstructures in 316L steel selective laser melting

Xin Zhou[a,c,*], Yuan Zhong[b], Zhijian Shen[a,b], Wei Liu[a]

[a] School of Materials Science and Engineering, Tsinghua University, 100084, Beijing, China
[b] Department of Materials and Environmental Chemistry, Arrhenius Laboratory, Stockholm University, S-106 91, Stockholm, Sweden
[c] Science and Technology on Plasma Dynamics Laboratory, 710038, Xi'an, China

**Corresponding authors:** Xin Zhou, zhouxin12@tsinghua.org.cn

## Abstract

The unique sub-grain patterns have been found in some particular alloys (316L, Al-Si, Co-Cr-Mo) selective laser melting (SLM), the submicron-scale cellular, elongated cellular or even band structures are always coexisting inside one single macro-solidified grain. Furthermore, the cellular structures are symmetrical with hexagonal, pentagonal and square cellular patterns where the cellular size is only around 1μm. Single-layer and bulk 316L SLM experiments are presented that reveals the forming mechanism of these sub-grain cellular microstructures. Complex cellular sub-micron patterns were formed by the local convection and Bénard Instabilities in front of the solid/liquid (S/L) interface (so-called mushy zones) affected by intricate temperature and surface tension gradients. In other words, this nonlinear self-organization phenomenon (Bénard Instability) occurring at the S/L interface is superimposed on the macro-grain solidification process to form the sub-grain patterns/structures and elemental microsegregations. This simple and unified explanation can be expanded to other eutectic alloys formed by SLM, like the Al-Si system.

## 1. Introduction

Selective laser melting (SLM) is a member of the additive manufacturing family of technologies whereby a three-dimensional (3D) part is built layer by layer by laser scanning of a precursor powder bed [1-4]. The physical feature of SLM is very similar to the micro-beam laser welding; a high energy and fast scanning micro laser beam (~100 μm) induces small melt volumes or melt pools which then solidify rapidly. The non-equilibrium solidification and multiscale hierarchical SLM microstructures will respond differently than conventional processing technologies [5, 6].

Unique sub-grain cellular and/or band morphologies are observed as typical features achieved by SLM in a range of metals or alloys and this might provide new routes for tailoring metal properties and performances [2, 7, 8]. A sub-grain (0.5 μm) cellular structure found inside each individual large grain in 316L SLM where reported by Saeidi [9]. Molybdenum was found to be enriched at the sub-grain boundaries, see Fig. 1a. Similar austenitic cellular colonies with sub-micro scale primary cell spacing in 316L SLM was reported by Yadroitsev [10]. A Mo-rich area was found in the central region of the track, see Fig. 1b. The latter author also reported the formation of colonies of coherent cells in AISI 420 austenitic stainless steel by SLM, in Fig. 1c [11]. The microstructure of 1Cr18Ni9Ti steel by SLM was mainly composed of cellular dendrites and demonstrated that the dendrite spacing increased with increasing powder layer thickness by Ma [12], see Fig. 1d. The resulting cellular/dendritic morphologies in SLM of 18Ni300 maraging steel contains very fine solidification cellular grains, with intercellular spacing smaller than 1μm, as reported by Casalino [13]. Here the differences between larger solidified macro-grains and fine sub-grain cellular structures/patterns should be emphasized. Firstly, the sub-grain structures are found to be less or around 1μm while the coexisting macro-grains are around tenths of micrometers. Secondly, two adjacent macro-grains always have different crystallographic orientations or high-angle grain boundaries while the sub-grain structures (within one macro-grain) have the same orientations. Thirdly, and most important, the sub-grain structures in a single macro-grain could have variable patterns (cellular, elongated cellular and even bands/stripes), which indicate a unique forming mechanism.

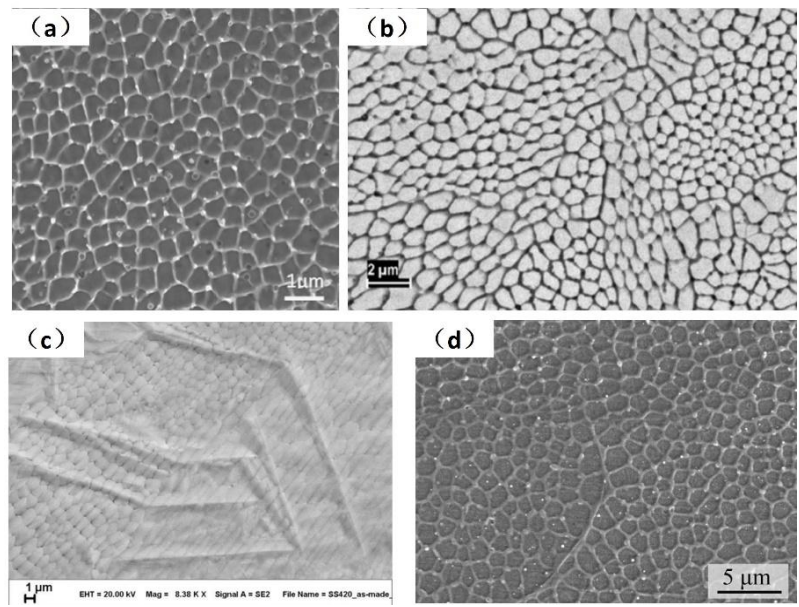

Fig. 1. Fine sub-grain cellular structures are shown by SEM images of SLM steel; (a) 316L [9], (b) 316L [10], (c) AISI 420 [11], (d) 1Cr18Ni9Ti [12]. Only (c) is exposed at the as-melted surface, whereas other images are from mechanical polished and chemically etched surfaces.

Besides austenitic stainless steel, Al-Si is another alloy system which can form the unique sub-grain cellular microstructures. Bartkowiak performed SLM on Al-Si, Al-Cu and Al-Zn powder systems [14]. Cellular structures within very fine Si precipitations were found in the Al-Si system, but in other systems (Al-Cu, Al-Zn) these were not found, see Al-Si in Fig.2a. Dinda also found equiaxed dendrites in an Al matrix with a network of Si particles and the dendrite arm spacing was 1.4~1.7 μm [15]. Kempen found very fine sub-grain microstructures and fine distributions of Si phase in AlSi10Mg SLM parts [16]. This contributed to a higher hardness and strengths of the alloy; the microstructure is seen in Fig.2b. Using the same alloy composition and SLM, a very fine cellular–dendritic solidification structure with a size smaller than 1μm was observed by Thijs [17]. The grey cellular features in the SEM images were primary Al metal with small, distributed fibrous Si particles (white), as seen in Fig. 2c. Other similar results of SLM of Al-Si alloys can be found in Refs. [18-21], which prove that fine supersaturated Al-rich cellular structures along with Si at the boundaries are very common, see Fig. 2d.

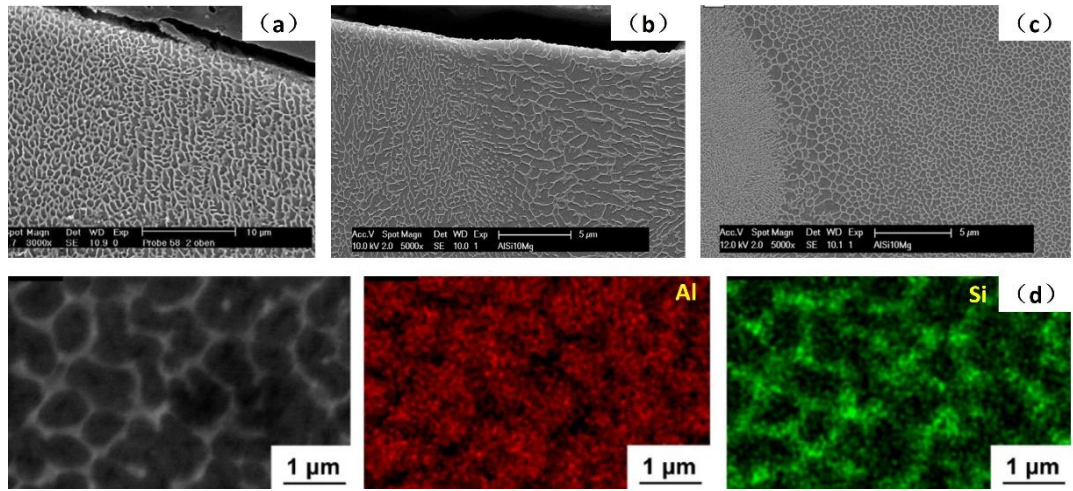

Fig. 2. SEM images of the SLM Al-Si alloy where the small cellular Al-matrix has very fine Si precipitations in (a) [14], (b) [16] and (c) [17]. SEM micrograph with an EDX composition map (enhancing Si) is seen in (d) [19].

Besides austenitic steel and the Al-Si system, the sub-grain cellular morphologies can also be found in some other alloys processed by SLM; they are characterized by FCC structures or eutectic structures. Thus, Carter reported fine cellular sub-grain structures in Ni625 (Ni-Cr21.5-Mo9-Nb3.6) by SLM [22]. Here large amounts of Nb and Mo concentrated at cellular boundaries. In addition, Hedberg and Qian reported the use of SLM for Co-Cr-Mo compositions that gave fine cellular and elongated cellular microstructural features with cell boundaries enriched in Mo (Co depleted) [23]. The same cellular structures were observed in a SLM Co-Cr-Mo sample, at the transverse cross section normal to the building direction, by Takaichi [24]. Song fabricated NiCr alloy parts by SLM and found unusual fine columnar microstructural architectures [25]. Similar results are also reported in a CoCrW alloy by Lu [26]. These sub-grain cellular patterns, that tend to appear in some eutectic and FCC alloys by SLM, were discussed by Song in [6]. Song concluded that molten pool boundaries with fine cellular dendritic structures had been observed in 316L, 304 and Ni625 alloys, but these typical microstructures were not present in Ti-6Al-4V (HCP), Fe (BCC) and ferrous alloy metal matrix composites. This statement was supported by Vrancken that reported a Ti-6Al-4V(-ELI) SLM microstructure consisting of acicular martensitic α′ within prior columnar β grains (50-150 μm) changed solidification mechanism when mixed with 10 wt.% Mo powder [27]. The new metal-metal composite (Ti-6Al-4V+10Mo) changed from planar to cellular mode and the new cellular β grains (5-15 μm) were significant smaller than the earlier columnar grains. More important, within each β grain a cellular substructure with an intercellular spacing of less than 1μm was present and microsegregation of the elements Mo, Al and V took place.

The forming mechanism of sub-grain cellular microstructural features in SLM (and other laser surface melting technologies) is still an open question, although some researchers have tried to explain it. The given explanations can be classified into three classes, where the first is involving compositional fluctuations and the constitutional supercooling theory. The ratio of temperature gradients and growth rates (G/R) decides the grain growth morphologies; heat accumulation may also provide the opportunity for a transition from columnar to equiaxed structure (CET) [9, 10, 13, 28-36]. The second class of explanation is the interface stability theory; the formed microstructure is related to the changes in velocity of the solidification front, rapid acceleration, oscillatory morphological instability and nonequilibrium trapping of solute [37-43]. The third class of explanation is either about the surface tension driven convection or a diffusive transport of impurities or of one of the constituents of the material [44-46]. Qu systematically investigated the solid-liquid interfacial morphology evolutions of Al-1.5%Cu in rapid solidification [47]. Qu found that the dendrite tip shape was an important parameter affecting the dendrite to cell transition. The transition can be described by specific linked conditions of velocities, temperature gradients and alloy compositions.

All these explanations seem logical, but none of them can be considered as a unified explanation for all observed cases. The SLM process itself raises some specific conditions; firstly, the size of a SLM melt pool is limited, width and depth are around 100~200 μm, and the process can be treated as a thin film flow in which the gravity effect can be ignored. Secondly, the melt pool has extreme temperature gradients (~10 000 K/mm); inevitable giving intense convection and turbulent heat- and mass-transfers [48]. Based upon these realities, the focus of this study will be thermocapillary (Marangoni) or solutocapillary flow in the SLM melt pool. It is known to play an important role in hydrodynamics, heat/mass transfer and solidification microstructure formation in melt alloys systems [49]. In addition, the hexagonal cellular convective pattern (Bénard cell) caused by surface-tension-driven instability (Bénard-Marangoni-Instability, BMI) is also an important physical phenomenon, as shown in Fig. 3a [50, 51]. These cellular convections in a melt pool have very important impact on the crystal growth, because the growing (moving) crystal interface can react to flow oscillations and can incorporate them as solidification microstructures [36, 51-53].

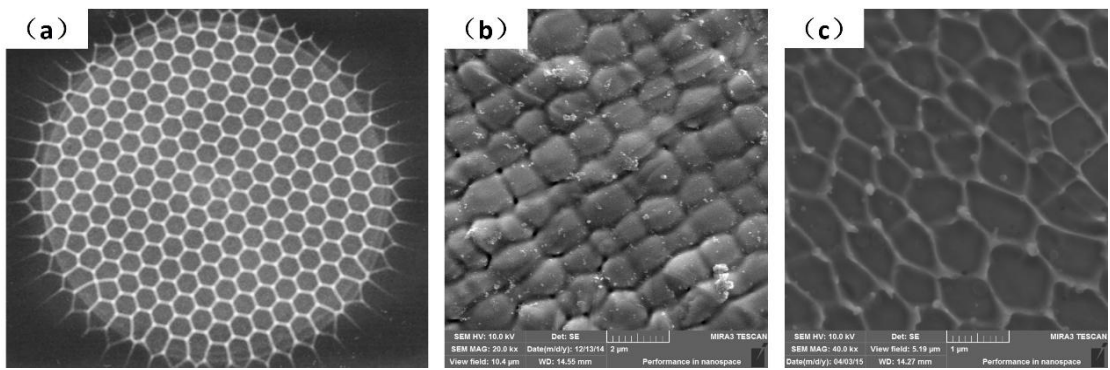

Fig. 3. Hexagonal convection patterns of the Bénard-Marangoni-Instability (a) [51]; as-melted top surface morphologies of 316L steel by SLM (b), etched fine sub-grain cellular microstructures of 316L steel by SLM (c). Comparing the similar size of cellular microstructures in (b) and (c)

On the basis of melt pool convection and Bénard-Marangoni-Instability theory, single-layer and bulk 316L by SLM experiments were conducted in this study to explain the mechanism of sub-grain cellular microstructures. The relations between surface-tension-driven Bénard conventions, SLM as-melted top surface morphologies and the fine sub-grain cellular microstructures are also considered as illustrated in Fig. 3a, 3b and 3c, respectively. There are reasons to believe that complex cellular sub-grain patterns are formed by the local convection and Bénard Instability in front of the solid/liquid (S/L) interface (so-called mushy zone) and these are affected by extreme and intricate temperature gradients. This scenario can also be considered as a nonlinear self-organization phenomenon, which is superimposed on macro-grain solidification to form the sub-grain patterns and microsegregations. This explanation seems reasonable and is unifying as it can be expanded to other eutectic alloys prepared by SLM, *e.g.* the Al-Si systems.

## 2．Materials and methods

### 2.1 Material

316L stainless steel powder granules with an overall chemical composition of 17 wt.% Cr, 10.6 wt.% Ni, 2.3 wt.% Mo, 0.98 wt.% Mn, 0.4 wt.% Si, trace amounts of S, C, P, O, N and the balance being Fe is used as precursor, supplied by Sandvik Osprey Ltd., Neath, UK [9]. The powder granules are spherical with particle size of 22~53μm determined by a laser diffraction analyzer (Mastersizer 2000, Malvern Instruments, Worcestershire, UK). Before the experiments, the powders granules are sieved (50μm) under argon to reduce the agglomeration and to improve fluidity.

### 2.2 Experiment arrangement and procedures

All SLM experiments were conducted on a Renishaw AM250 facility equipped with a SPI redPOWER 200W ytterbium fiber laser, operating at 1,071nm wave length and 75 μm beam diameter ($\Phi_{99\%}$) (Renishaw AMPD, Stone, UK). The laser runs in modulated operation (pulsed with TTL trigger).

Single-layer SLM tests were performed to study the as-melted top surface morphologies, where the substrate was a rolled 316L plate, size of 10cm×8cm×3cm, and polished with 1000# abrasive paper. The polished surface was then coated with a black paint layer to reduce the reflection of laser energy. The powder layer was deposited on the black painted surface with a thickness of 50 μm by using the wiper system on the SLM machine, Fig. 4a. The laser beam then scanned the single powder layer with power 190W, scan speed 700 mm/s, line spacing 0.05 mm, and "zigzag" scan strategy. Oxygen content in the chamber was set as 1000 ppm (recommended by Renishaw). The single-layer powder appearance after an interrupted laser scan is shown in Fig. 4b.

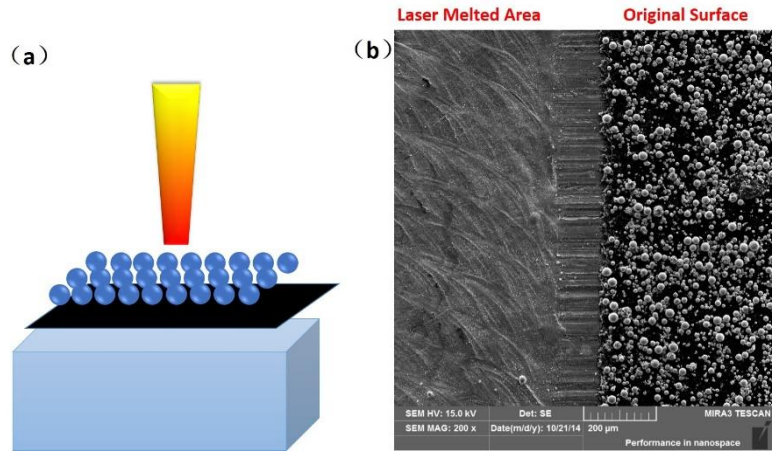

Fig. 4. An illustration of the single-layer powder SLM experiment (a). The top surface appearance after and before laser scanning is shown by a divided SEM image with left and right areas, respectively (b).

The 316L SLM bulks were fabricated using the same laser parameters as in the single-layer SLM experiments, except line spacing was 0.125 mm. The scan strategy was "cross hatching" which had long bi-directional scanning vectors and performed 67° angle rotation of scanning direction between adjacent layers.

## 2.3 Microstructural characterization

Samples were microscopically characterized using a TESCAN MIRA 3LMH scanning electron microscope (SEM) from TESCAN (Brno, Czech Republic). The SEM samples were ground using sand paper in a Buehler abrasive belt grinder and followed by polishing with a set of decreasing diamond size suspensions with a final 1μm size. Chemically etching occurred in an acidic water solution containing 2% HF - 8% $HNO_3$ - deionized water for, 10 minutes at 25℃. Transmission electron microscope (TEM) tests were performed on a JEM-2100 (JEOL, Tokyo, Japan). The TEM samples were first ground down to 70 μm thickness and then twin-jet electropolished to electron transparency; electrolyte was 10% perchloric acid in methanol and the temperature was maintained at -30℃.

## 3. Results

### 3.1 The top surface morphologies of single-layer laser melting

Clear melt tracks and surface ripples can be observed at the single-layer laser melting surface, as seen in Fig. 5a. These features are induced by the melt pool dynamics/oscillations and are typical surface phenomena observed in both welding and SLM [54, 55]. Other solidification patterns can be found, as cell dendrites (elongated cells and strips) with different directions, but the main direction is always pointing to the melt pool center, see Fig. 5b. The solidification patterns in Fig. 5 c-d actually represent the melt flow direction during the laser melting process, from the laser melt pool edge to the center. Three particular patterns can be distinguished by SEM at 5000X magnification; the first mode shows the mixed stationary hexagonal, pentagonal and square cellular patterns where the cellular size is only around 1μm, see Fig. 6 a-b and Fig.7. The second mode displays a "drifting cell", which can be considered as an elongated hexagonal cell, see Fig. 6 c-d. Finally, the third mode shows the appearance of "long strips" that have a width of only 1 μm, but with a length over 50 μm, see Fig. 6 e-f. The forming mechanism of these three as-melted top surface patterns can be explained by the interaction between convection instabilities in front of the solidification front and solute transport behavior, which will be further discussed in Section 4 below.

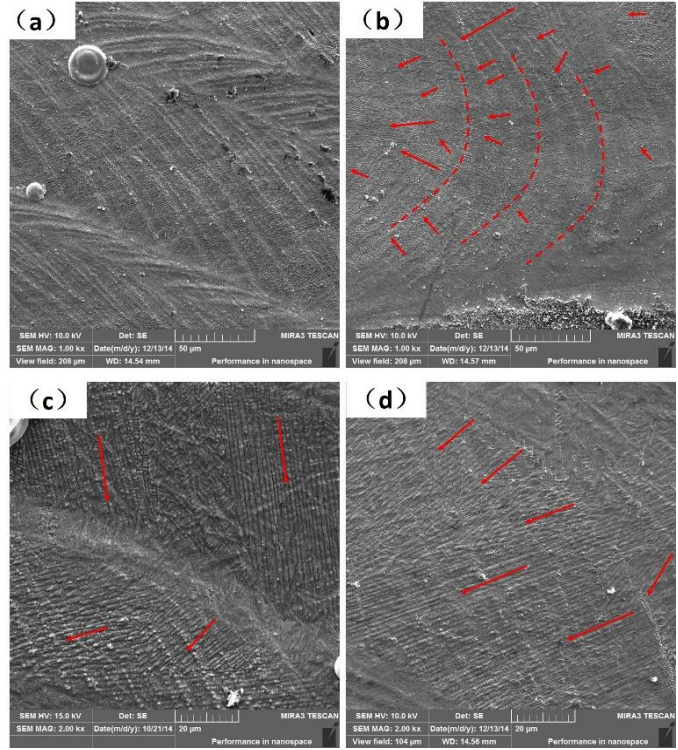

Fig. 5. Top views of a single-layer powder surface after laser melting observed at SEM with 1000X (a, b) and 2000X (c, d) magnification. The red dashed lines demonstrate the laser melt pool boundaries; the red arrows demonstrate the complex melt flow directions.

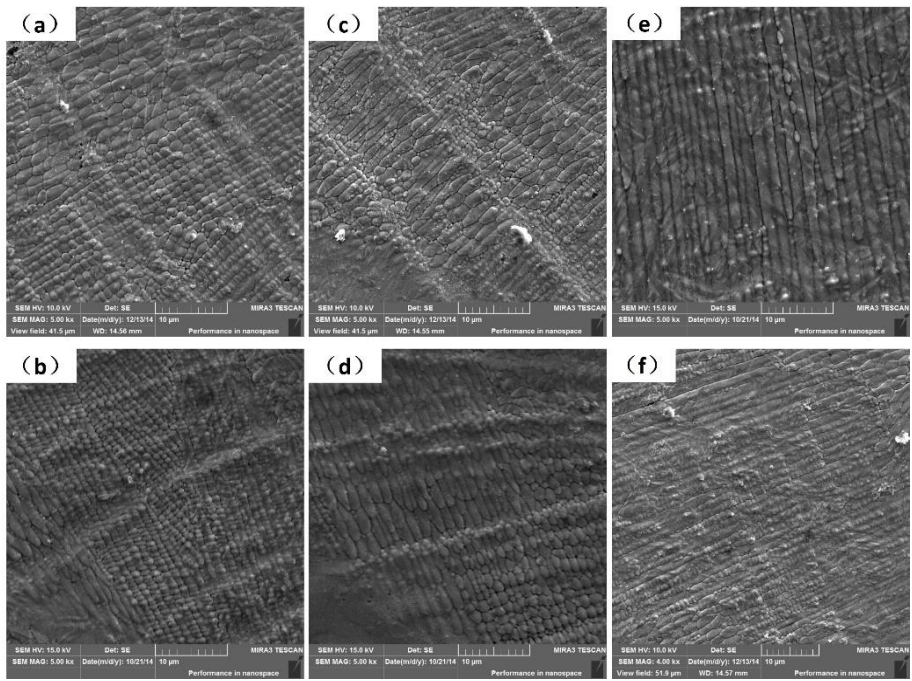

Fig. 6. Different solidification morphologies of the top surface of a single-layer powder after laser melting. Mixed hexagonal, pentagonal and square cellular patterns are shown in (a, b),

elongated drifting cellular patterns in (c, d) and strip flow patterns in (e, f).

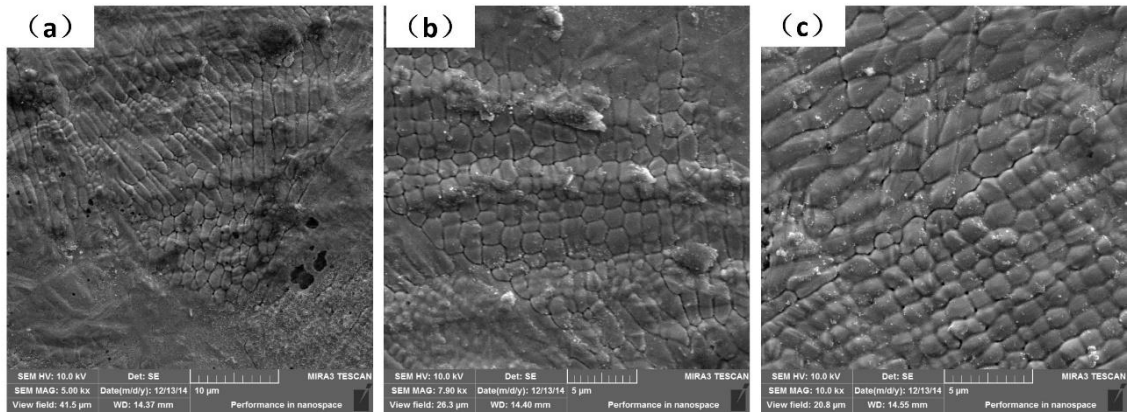

Fig. 7. SEM images showing the geometry patterns in detail, demonstrating the mixed hexagonal, pentagonal and square cellular pattern in the images (a) to (c), respectively.

**3.2 The sub-grain cellular/bands microstructures**

The solidified macro-grain boundaries (classified by grain orientations) in SLM 316L stainless steel are shown in Fig. 8. The size distributions of these irregular macro-grains are not uniform; there are both some larger grains with size over 50 μm and some smaller grains, but the median macro-grain size according to the graphical analysis results is of the order 10 μm.

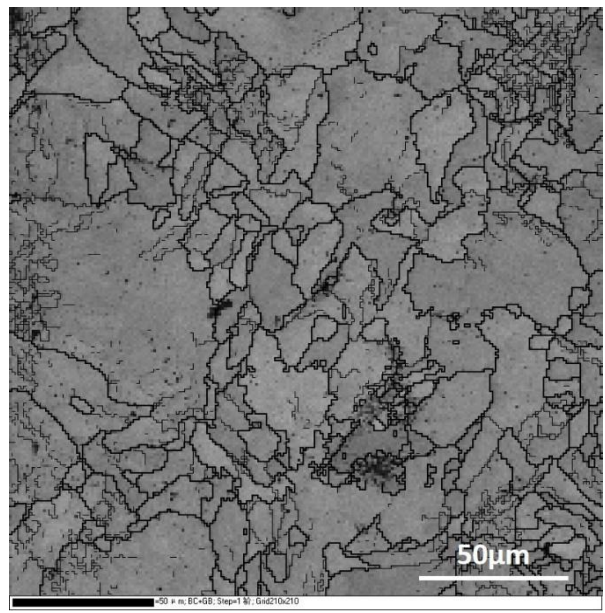

Fig. 8. The high-angle grain boundaries were analyzed by EBSD and presented at lower

magnification.

The macro-grains have very complex substructures and SEM images from polished and chemically etched surfaces are given as examples in Fig. 9. Distinct and complex fine band/cellular sub-grain microstructures are revealed. There exist cellular and elongated cellular sub-grain patterns illustrated in Fig. 9 a-c. These are analogous in sizes and shapes to the as-melted top surface hexagonal patterns shown in Fig. 6 and 7. At the same time, the band structures in Fig. 9 d-f are analogous to the as-melted top surface strip patterns shown in Fig. 6 e-f. Regular cellular and elongated cellular structures appear simultaneously, as highlighted in Fig. 9 b. This demonstrates that the transitions from regular cells to elongated cells are natural under some circumstances. Furthermore, transition from cellular to strip patterns can also be found in Fig. 9 e. The band structures are formed in the following sequence: regular cells → elongated cells → bands. These transitions are very confusing as they exist in one single macro-grain. Current theoretical explanations of constitutional supercooling, columnar to equiaxed transition (CET) and lateral instabilities beneficial to the growth of secondary arms are all not very convincing. Therefore we suggest that these observations are nonlinear self-organization phenomena under the strong marangoni convection in front of the S/L interface, as proved and discussed in detail below.

Details of sub-grain cellular and band structures exposed by SEM upon the transverse cross section (normal to the building direction) are presented in Fig. 10. The two clusters of bands in Fig. 10a are not solely by an epitaxial growth mechanism. The sub-grain boundaries are obviously more resistant than the interior of sub-grains toward the etching acidic media. The EDS line scan analysis confirm micro-segregation and concentration of elements that are more corrosion resistant into the boundaries, see Fig. 10c.

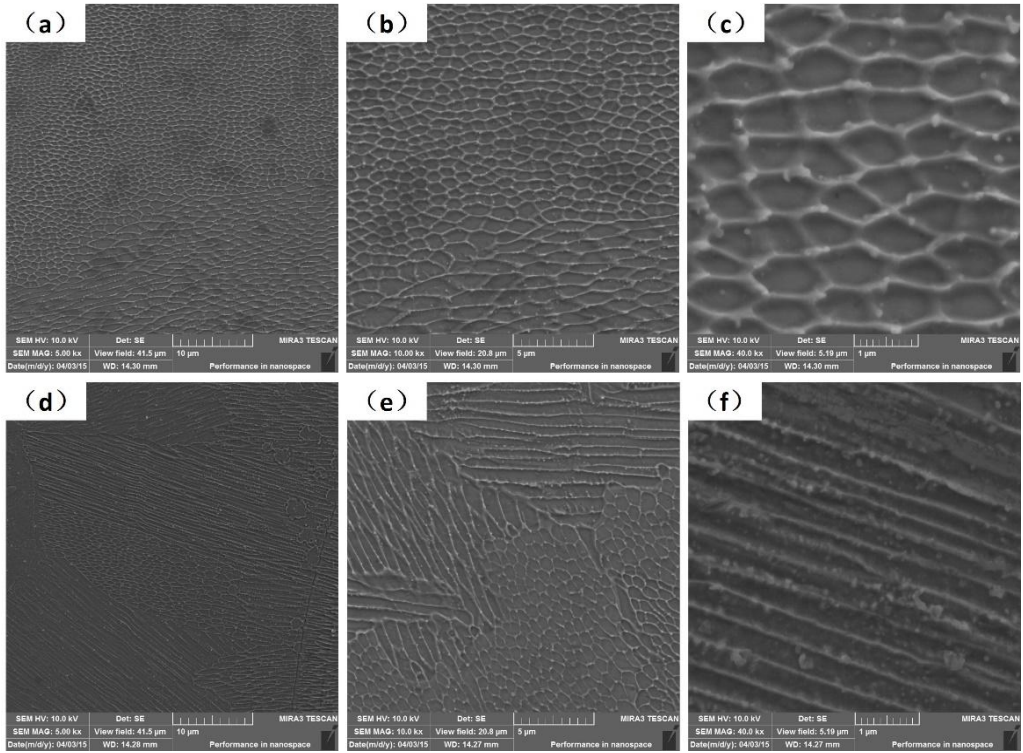

Fig. 9. The SEM images of a variety of sub-grain microstructures; with mixed cellular and band morphologies (a-f). All the images were exposed upon a transverse cross-section normal to the building direction.

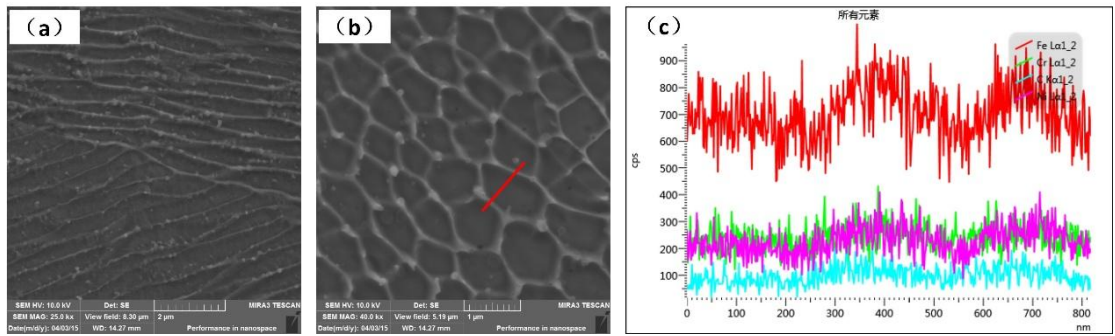

Fig. 10. Fine band (a) and cellular (b) microstructures are seen with a size around 0.5~1μm. The EDS line scan analysis (c) shows element distribution along the line shown in (b). The variations of Fe, Cr, Ni contents are obvious between the boundary regions and the interior of the sub-grains, cf. text.

### 3.3 The TEM observation

Sub-grain bands and cellular microstructures can be distinguished also by TEM images, as seen in Fig. 11. The black circular inclusion in Fig. 11a is an amorphous Cr-Si-O particle which has been discussed in [9]. Clearly dislocation tangles and dislocation cell structures are found in the TEM images. Occurrence of dislocation structures are normally achieved for plastically deformed steels, but do happen in SLM samples. Another observation is that the dislocations are not homogeneously distributed in the material, as in some area the density is very high while in other areas it is absent. This may related to the asymmetry of temperature gradients, impurity segregations and possible solid-state phase change of ferrite to austenite during rapid cooling.

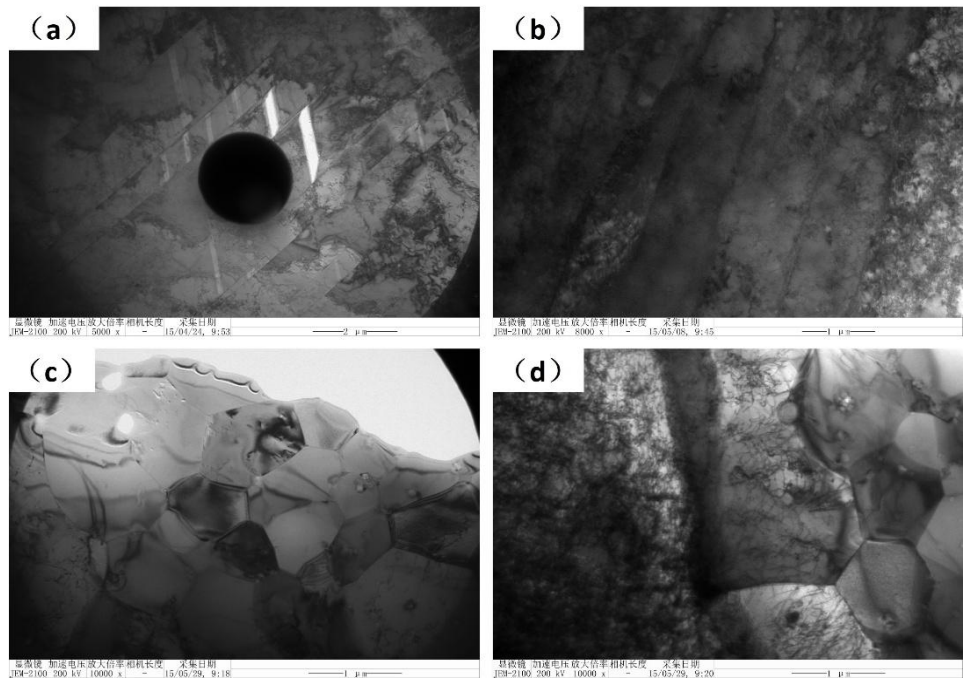

Fig. 11. TEM images are shown from band (a, b) and cellular microstructures (c, d). The black circular inclusion in (a) is a Cr-Si-O particle.

## 4．Discussion

### 4.1 The complex thermocapillary and solutocapillary convections in the melt pool

The SLM technique or micro beam laser welding comprises very complex physical processes. The liquid metal is affected by heat/solute transport and at the same time intense convective vortex movements driven by asymmetrical temperature gradients, surface tension gradients and laser recoil forces (keyhole). Results from finite element simulation of the SLM process used on 316L steel can be seen in Fig 12-14 [56]. A finger shaped melt pool forms at ultra-high heating/cooling rates when laser irradiates the surface in a straight line. The node temperature increases rapidly to a maximum around 2400K in Fig. 12. After the laser moved away the node temperature drops rapidly and the cooling rate can be estimated to around $6 \times 10^4$ K/s (G R, K/cm cm/s), see Fig. 12b. The temperature gradients in different cross profiles are found in Fig. 13. The width of the elongated melt track is around 150 μm, but the length is near 1mm. A high temperature gradient exists at the melt track edge perpendicular to laser moving direction, in Fig. 13a. The value of $G_x$ can be calculated as high as $1.3 \times 10^4$ K/cm, see Fig. 13b. A longitudinal-section view reveals that the melt has a maximum depth of 200 μm and a tail with gradual reduced depth, in Fig. 13c. The temperature gradient $G_y$ along the tail (laser moving direction) is only $1.0 \times 10^3$ K/cm, see Fig. 13d. At a cross-section view, the temperature gradient $G_z$ in the melt pool bottom (perpendicular to laser scan surface) is about $6.2 \times 10^3$ K/cm, in Fig.13 e-f.

The mentioned non-uniform temperature gradients found in the melt pool can induce surface tension variations and generate thermocapillary flow [57]. The direction of a thermocapillary flow depends on the temperature coefficient of the surface tension, $\partial \sigma / \partial T$. For a metallic melt having negative $\partial \sigma / \partial T$; the higher surface tension of liquid metal near the edge (cooler) and the thermal-capillary force induced by the surface tension gradients pull the liquid metal away from the center. Therefore the melt will flow from the center to the edge (so-called Marangoni flow), as plotted in Fig. 14b. When the flow approaches the edge it sinks and reverses along the bottom to the center where it will rise to the surface again; a full circulation loop. A convective vortex movement is formed, as illustrated in Fig. 14. In addition, inhomogeneity of surface tensions may also result from temperature or concentration variations of other solutes altering the surface energies, *e.g.* dissolved surface-tension active components (S, O) and element enrichments of a multi-element alloy. Surface tension will be a function of temperature *T* and the contents of other elements [58]:

$$\frac{\partial \sigma}{\partial T} = -A - R\Gamma_s \ln(1 + K_{seg} a_i) - \frac{K_{seg} a_i}{1 + K_{seg} a_i} \cdot \frac{\Gamma_s \Delta H^0}{T} \qquad (1)$$

Where $\partial \sigma / \partial T$ is coefficient (temperature and surface-active elements) of surface tension;

$R$ is the gas constant; $\Gamma_s$ is the saturated surface excess; $K_{seq}$ is the equilibrium absorption coefficient of surface-active elements; $a_i$ is the activity of surface-active elements (weight %); and $\Delta H^0$ is the standard heat of adsorption. From *Equ.1*, all observed structural features can follow from changes of the local surface tension in front of the S/L interface and the thermocapillary flow mode. Two examples are the rejected elements of Si, Cr, Mo during austenitic solidification in 316L and the precipitation of Si phase in the Al-Si eutectic system, these elements can change the local surface tension gradients.

In our experiments, although the SLM chamber environment is carefully controlled, it is still difficult remove all surface-active elements completely, *e.g.* oxygen. Some residual oxygen is present in the precursor powder and in the SLM machine chamber (＜1000 ppm). In the SLM process; the center of laser melt pool has the highest temperature and as $\partial\sigma/\partial T$ is a negative value the melt at the top surface flows outward (*Equ.1*). In the edge area, with significant lower temperature, the $\partial\sigma/\partial T$ changes to a positive value and the melt flow inverts inward from the edge to the center. The SEM images presented in Fig. 5 can be understood by this mechanism. Firstly, at the melt pool edge the largest temperature gradients exist with a dissolved surface-active element film, resulting in the described melt flow from edge to the center. Secondly, at an invariable cooling rate of $6\times10^4$ K/s (G R) and the smallest temperature gradient being along the tail ($1.0\times10^3$ K/cm), the growth rate R has a maximum value of 600 mm/s (laser speed 700 mm/s) from the tail to the center. Most important is that these ultra-high, nonlinear and asymmetrical temperature gradients can initiate intense melt jets and even turbulence instabilities, with surface flow rates higher than 1000 mm/s, which will then form complicated flow patterns and solidification microstructures [59].

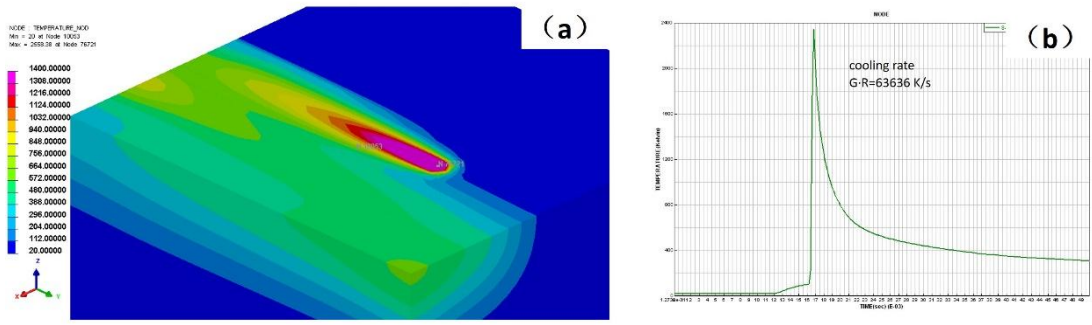

Fig. 12. FEM results of 316L by SLM; isometric view (a) and plot of node temperature vs. time relationship (b). The cooling rate (G R, K/s) can be calculated.

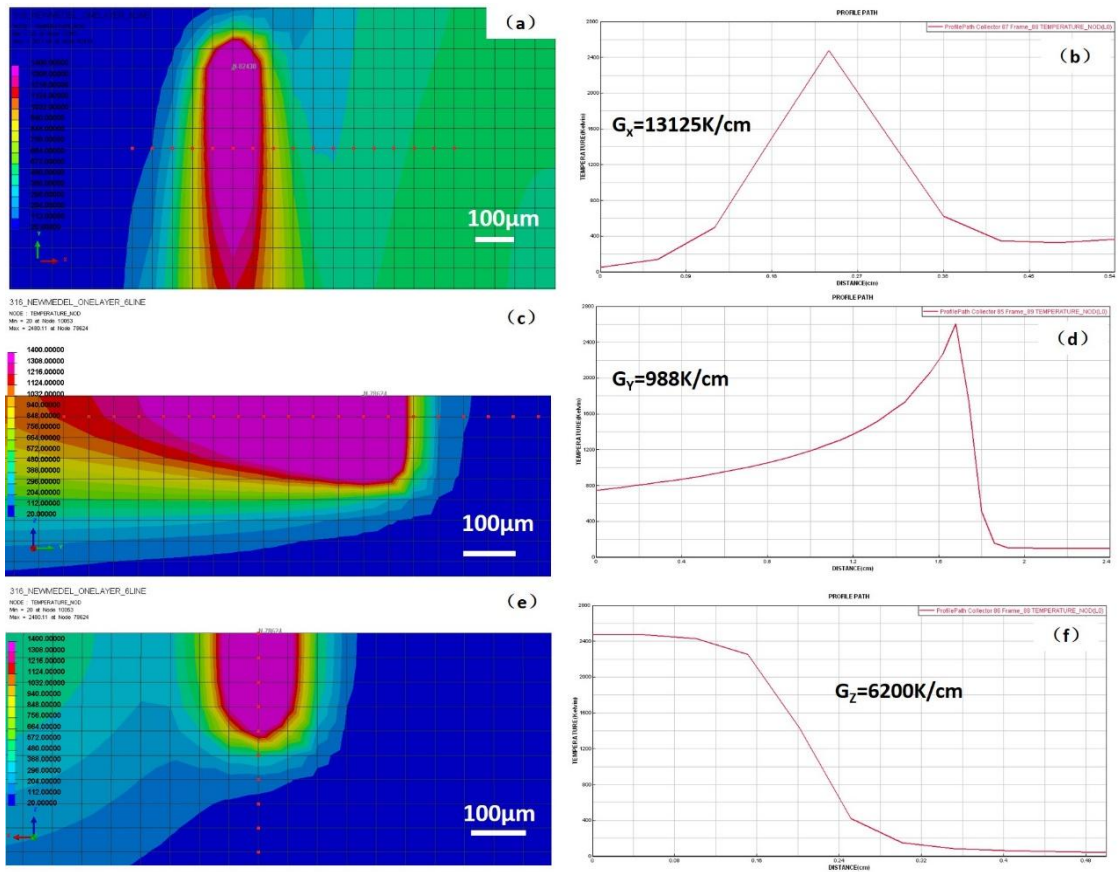

Fig. 13. FEM results of 316L by SLM. The melt pool by a top view (a), a longitudinal-section view (c) and a cross-section view (e), where a plot of node temperature vs. distance can be found in (b), (d) and (f), respectively. The involved nodes are also marked separately in (a), (c) and (e). Temperature gradients of the three different directions ($G_x$, $G_y$, $G_z$) can be calculated, where Gx represents the temperature gradient in the top surface melt pool edge, Gy represent the gradients in the melt pool tail and Gz represent the gradients in the melt pool bottom.

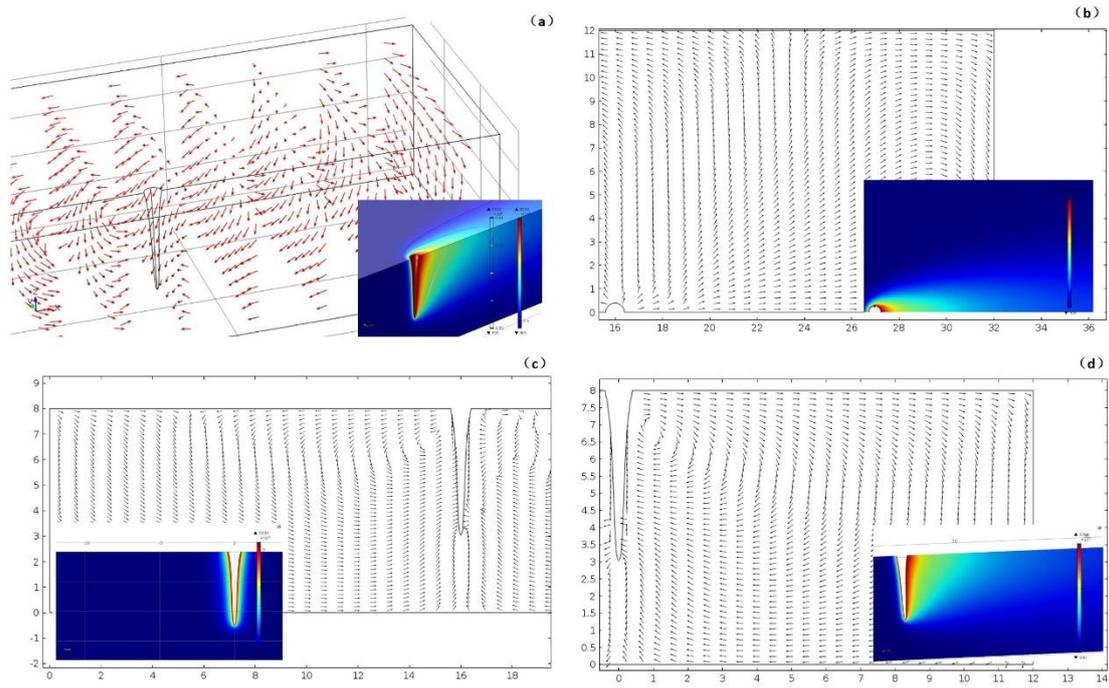

Fig. 14. Flow field computation results (with keyhole) of a melt pool formed by 316L by SLM; an isometric view (a), a top view (b), a cross-section view (c) and a longitudinal-section view (d). The corresponding temperature fields are also plotted.

### 4.2 Bénard instabilities and cellular microstructures in front of the S/L interface

For small melt volumes, rapid heating/cooling rates and limited solute redistributions, the solidification by SLM can be considered as a process of "no solid diffusion and limited liquid diffusion" according to classical solidification theory. The solute rejected by the growing solid front forms a solute-rich boundary layer (mushy zone) ahead of the growing front. Convection induced mixing in the liquid has a very important effect on solute segregation and subsequently formed solid microstructures. The temperature gradients are asymmetrical and inclined to the free surface, see Fig. 13. Coexistences of both vertical and horizontal gradients can be considered [60]. The vertical Marangoni number and horizontal Marangoni number are defined:

$$\mathrm{Ma}_{ver} = \left|\frac{\partial \sigma}{\partial T}\right| \cdot \Delta T_{ver} \cdot d \cdot \eta^{-1} \cdot \chi^{-1} \tag{2}$$

$$\mathrm{Ma}_{hor} = \left|\frac{\partial \sigma}{\partial T}\right| \cdot grad_x T \cdot d^2 \cdot \eta^{-1} \cdot \chi^{-1} \tag{3}$$

Where $\Delta T_{ver}$ is the vertical temperature gradient; $grad_x T$ is the horizontal temperature gradient; $d$ is the layer thickness of the instability ($1 \times 10^{-6}$ m in vertical, $75 \times 10^{-6}$ m in horizontal, as discussed below); $\eta$ is the dynamic viscosity (6.44 mPa·s); $\chi$ is the thermal diffusivity ($1.89 \cdot 10^{-5}$ m$^2$s$^{-1}$); $\partial\sigma/\partial T$= -0.39 mNm$^{-1}$K$^{-1}$. Thus, the vertical Marangoni number $Ma_{ver}$ (near the bottom and in front of a S/L interface) can be estimated roughly as 2000 and the horizontal $Ma_{hor}$ (near the track edge) is roughly about 180.

It is known that when a liquid layer has a vertical temperature gradient and the $Ma_{ver}$ becomes higher than the critical one ($\Delta T$ becomes higher than the critical $\Delta T_c$), the so-called Bénard-Marangoni-Instability occurs in the form of hexagonal cells driven by surface tension [51, 61-64]. The $Ma_{ver}$ and $Ma_{hor}$ calculated above are remarkable large and an instability is inevitable. Flow pattern maps can be found in Fig. 15. When $Ma_{ver}$ is predominant and vertical melt flow has priority, the flow patterns have cellular structures (DC and SDC). When the $Ma_{hor}$ is predominant and horizontal melt flow has the priority, the flow patterns have roll/strip structures (LR and SLR). The vertical cells (vertical instability) can be generated in a very thin layer ($1 \times 10^{-6}$ m) but the horizontal rolls (horizontal instability) need a longer surface convective region. As illustrated earlier the cell spacing was around 1μm, but the strips has a length over 50 μm, see the Figs. 6 and 7. Transitions between different convective patterns occurring by changing the governing parameter $Ma$ have been proved [63]. Another similar case is the electron beam melting of high melting point metals; the beam heats the free surface of the melt and simultaneous radiative cooling is very significant, resulting in drifting cellular structures on the melt surface [60, 65].

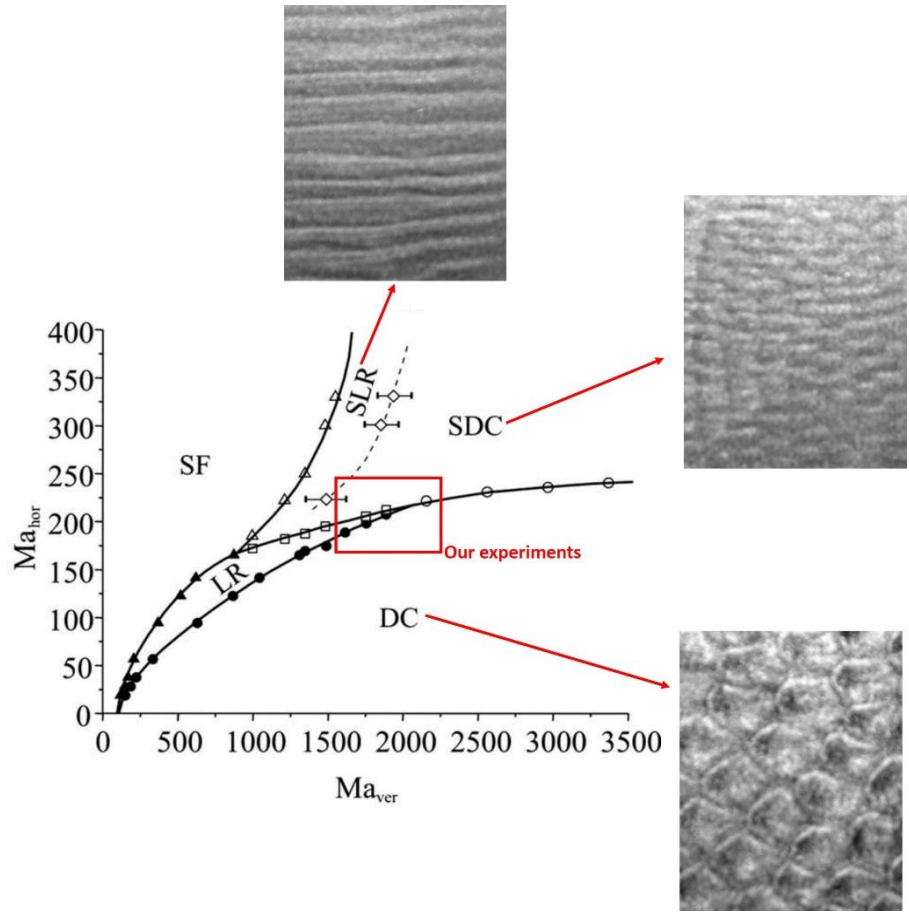

Fig. 15. Stability limits of thermocapillary flow and flow pattern transitions, as expressed by the relation between the vertical Marangoni number and horizontal Marangoni number [51, 60]. In the figure denotations are DC: Bénard-Marangoni drifting cells; SDC: surface drifting cells; LR: longitudinal roll; SLR: surface longitudinal roll; and SF: steady flow. The *Ma* of our experiments lies in the drawn red rectangle area, so all three patterns (SLR, DC, SDC) can be expected.

Based on the stability limits analysis and flow pattern transitions, the cellular sub-grain microstructure shown in Fig. 9c can be attributed to the Bénard-Marangoni-Instability. In a very thin layer ($1 \times 10^{-6}$ m) in front of the S/L interface and at the melt pool bottom, there exists a very strong temperature gradient shown in Fig. 16a [66]. Thereby Marangoni-Bénard convection with hexagonal structures can be generated in this thin layer and at the same time be superimposed on grain solidification, see Fig. 16b. In conclusion, the solidification mode and morphologies of macro-grains are controlled by G and R and the sub-grain patterns are controlled by flow instabilities and *Ma*; these two mechanisms are combined. Moreover, by the variations of fluid flow condition and the weld pool oscillation, the ideal hexagonal structures lose its stability to other geometries. Individual hexagons undergo local changes in topology and transform first into pentagons and then into squares; so the patterns within the macro-grain can be a mixture of hexagons, pentagons and squares, see Fig. 16c [67]. In addition, this Marangoni-Bénard instability can always be generated in front of the S/L interface, and advances accompany with the S/L interface movements until the sub-grain patterns are created in each of the macro-grains in the bulk.

In another situation, where a strong horizontal temperature gradient and horizontal fluid flow exists in the melt edge area, streak structures can be generated from melt pool edge to the center, as in Fig. 17a. These streak structures can also be superimposed on macro-grain solidification and the intragranular band patterns are generated, see Fig. 9f. Ideal conditions exist for structures seen in Fig.16 and Fig. 17a, but the actual temperature gradients in SLM pools are asymmetrical and complicated. These interactions can generate more complex sub-grain patterns, see Fig.17b. The cellular, elongated cellular and bands appear simultaneously and transitionally (Fig. 9). The as-melted top surface morphologies can also be explained by similar mechanisms and these morphologies are generated in the last stage of solidification, reflecting the complex surface Marangoni flows as plotted in Fig.18.

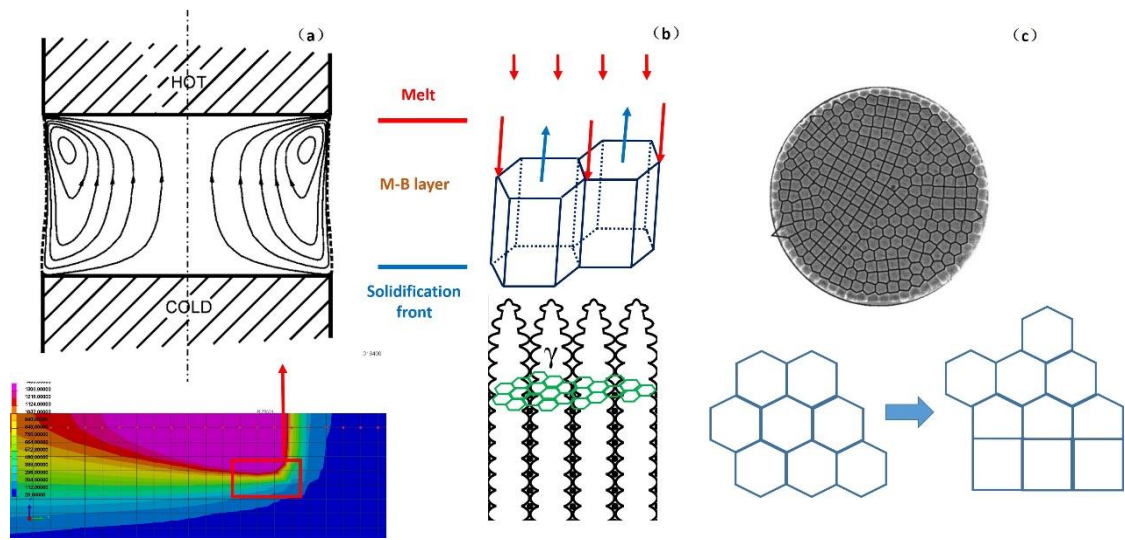

Fig. 16. Illustrations are given of the forming mechanism of cellular convective structures in front of a solid/liquid interface. (a) Flow structure of 2D steady thermocapillary flow, where all streamlines pass through a comparatively thin boundary layer [66]. (b) Marangoni-Bénard convection with honeycomb structures superimposed upon macro solidification. (c) With different Ma the transition from pure hexagonal pattern through the mixture of hexagons, pentagons to squares are also presented [67].

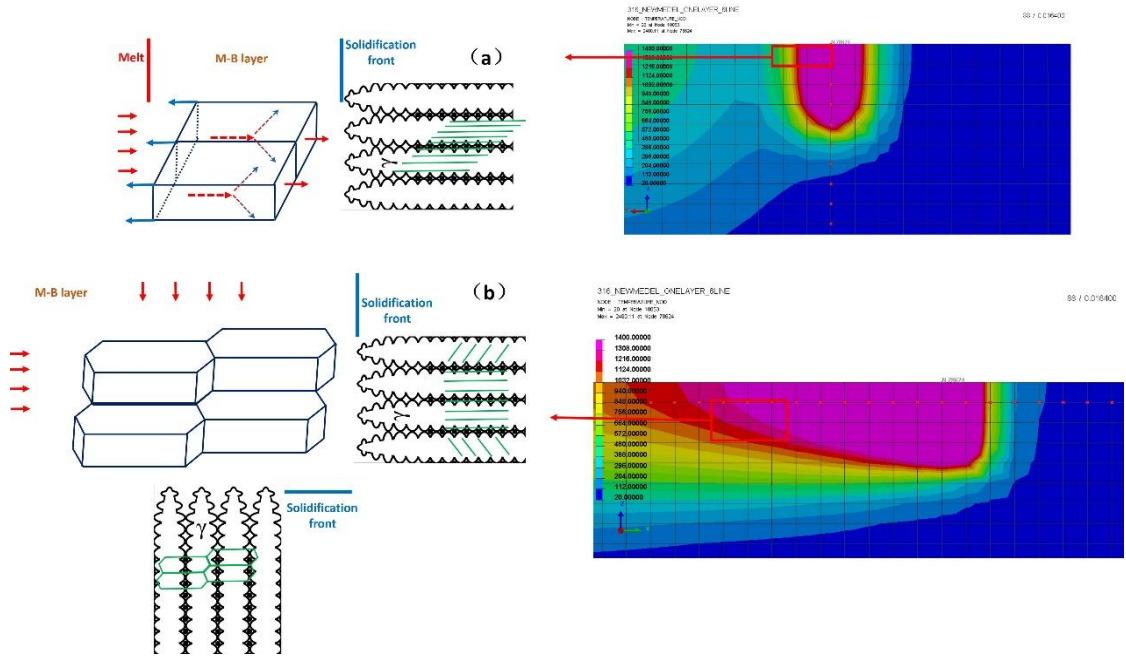

Fig. 17. The forming mechanism of the bands in (a) and mixed patterns with cellular, elongated cellular or band structures in (b).

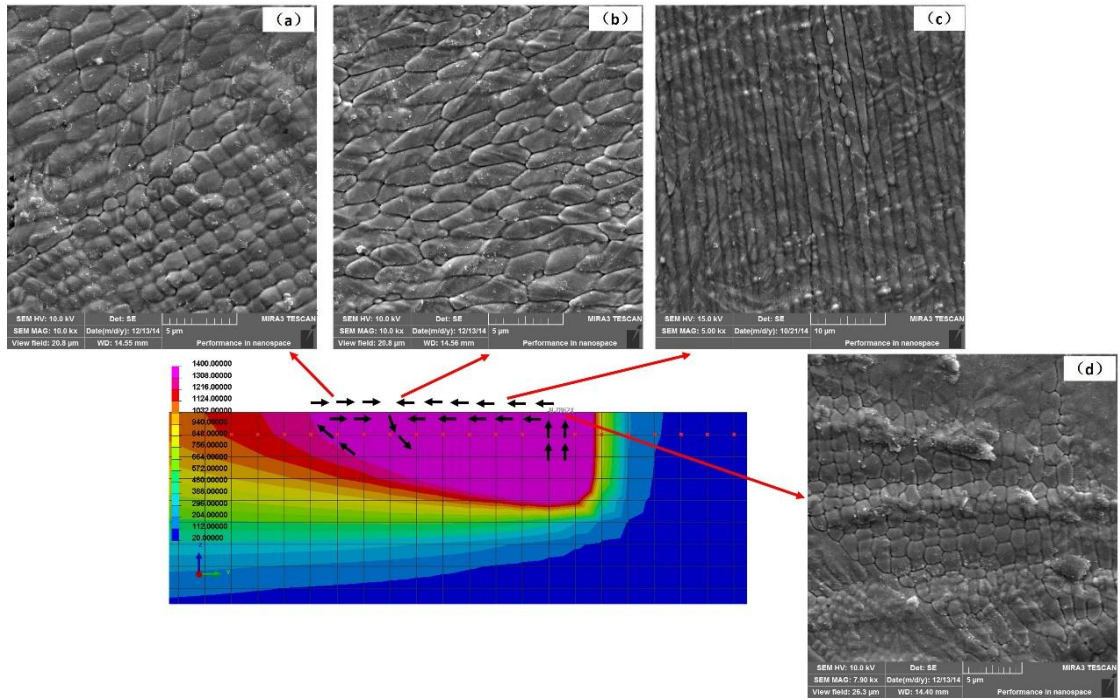

Fig. 18. The forming mechanism of as-melted top surface morphologies; (a) mixture of cellular and elongated cellular, (b) elongated cellular, (c) strips and (d) cellular structures.

### 4.3 The micro-segregations of sub-grain cellular structures

The thermocapillary flow and Bénard-Marangoni-Instabilities in front of the S/L interface can also be proved by the inclusions and micro-segregations in SLM sub-grain patterns of 316L and of a Al-Si alloy. The equivalents of Cr ($Cr_{eq}$) and Ni ($Ni_{eq}$) of the precursor powders, of the SLM cellular microstructure and of the Cr-Si-O inclusions are calculated according to Schaeffler predictive phase diagram, in Fig. 19c. For the precursor powder; it achieves 19.9 for $Cr_{eq}$ and 15.89 for $Ni_{eq}$, with $Cr_{eq}=$ *(Cr+Mo+1.5Si+0.5Nb)* and $Ni_{eq}=$ *(Ni+30C+0.5Mn)* [68]. For the SLM cellular microstructure; it attains 20.035 for $Cr_{eq}$ and 16.11 for $Ni_{eq}$. Finally, for Cr-Si-O inclusions it is 32.95 for $Cr_{eq}$ and is 15.84 for $Ni_{eq}$ (referred in [9]). The elemental composition of SLM cellular substructure is similar to the precursor powders, $Cr_{eq}/Ni_{eq}$ is 1.25 for powders and 1.24 for SLM sub-grain cells, *cf*. Fig. 19b. Comparing the Cr-Si-O inclusions with the precursor powders, austenite promoting elements (Ni, Mo) are reduced but ferrite promoting elements (Si, Cr, Ti) are increased, $Cr_{eq}/Ni_{eq}$ is 2.08 for Cr-Si-O inclusions, Fig.19a. According to the Schaeffler diagram and the WRC-1992 diagram, a single phase austenite will be formed by lower $Cr_{eq}/Ni_{eq}$ (＜1.37) and "austenite + acicular ferrite" will be generated for higher $Cr_{eq}/Ni_{eq}$ (≈2) [68, 69]. That means 316L steel solidifies with a single phase austenitic microstructure, which consumes the austenite-promoting Ni, and rejects the ferrite-promoting elements Cr, Si and Mo in the solidification front; thus the ratio of $Cr_{eq}/Ni_{eq}$ is increased ahead the S/L interface.

Another physical phenomenon named particle accumulation structures (PAS) will be introduced [66, 70-72]. PAS is the behavior of small particles of dilute concentration in a time-dependent (oscillatory) ring vortex thermocapillary flow. The evenly distributed particles will form clouds circulating in the vortex when affected by an oscillatory thermocapillary flow, see Fig.20a. Effect of PAS has an impact on the discussion of SLM, where the austenitic solidification of 316L steel will reject the ferrite-promoting elements Cr, Si and Mo. The two elements Cr and Mo increase the local surface tension (solutocapillary) and Si increases the melt liquidity. The enhanced capillary vortex will bring the rejected ferrite-promoting elements Cr, Si and Mo from the S/L interface to the instability layer. The elements Si and Cr have strong affinity to oxygen and will react with residual oxygen and form the mentioned Cr-Si-O inclusions. These will precipitate together with the Mo-enrichment at the cellular boundaries by Marangoni-Bénard convection and PAS mechanism. For the higher $Cr_{eq}/Ni_{eq}$ in the boundary, it changes the solidification mode from "full austenitic" to "austenite + acicular ferrite". The solid-state phase transformation from austenite to ferrite occurs during cooling and that is the main reason of high dislocation concentrations observed and it contribute to different resistances toward an acid etching agent at the sub-grain boundary.

These explanations can also be used for structures found in an Al-Si eutectic formed by SLM. The microstructure is significantly different when compared with a casted Al-Si alloy. By casting a continuous eutectic structure of Al and Si is displayed along with dispersed primary α-Al, see Fig. 20b [19]. The Al-Si SLM microstructure consists of cellular morphologies and EDS analysis reveals that Si is preferentially located at the cellular boundaries (thickness of about 200 nm) and consequently the cells of 500-1000 nm size is richer in Al, see in Fig. 20 c. Based on the Al-Si phase diagram, the solubility of Si in Al is 1.65 wt% at 850K but decreases to 0.06 wt% at 573K [18]. Thus, the solidifying front rejects Si ahead of the S/L interface. Under the mentioned Marangoni-Bénard convection and PAS mechanism, the rejected Si particles will be redistributed by the intense capillary flow. Similar patterns as seen for 316L SLM (cellular, elongated cellular, bands) are preferred and residual Si micro-segregations will be seen along such boundaries. This mechanism can also be tested by the results of different base plate heating's in Fig. 20 c-d [20]. The microstructures obtained with base plate heating and without base plate heating have some differences; the width of dendrites is increased and some laminar eutectic appear with base plate heating. That might be due to the reduced $\Delta T$ and driving force of convections giving a tendency to solidify more as an eutectic casting.

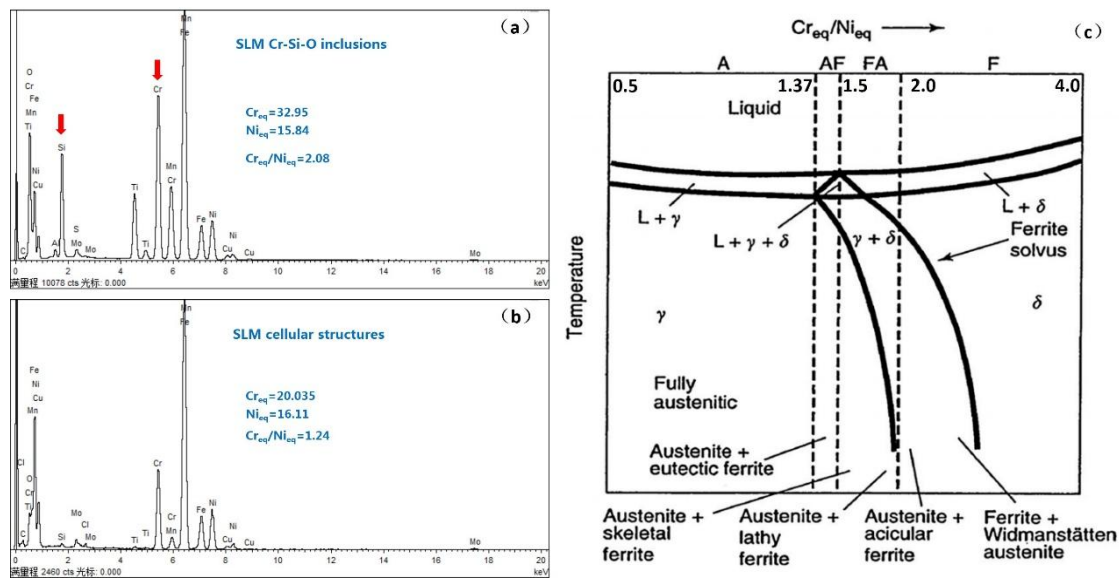

Fig. 19. EDS elemental analysis were done on a SLM Cr-Si-O inclusion (a) and a sub-grain cellular microstructure (b), resulting in calculated equivalent amounts of Cr and Ni. The Schaeffler diagram is shown in (c).

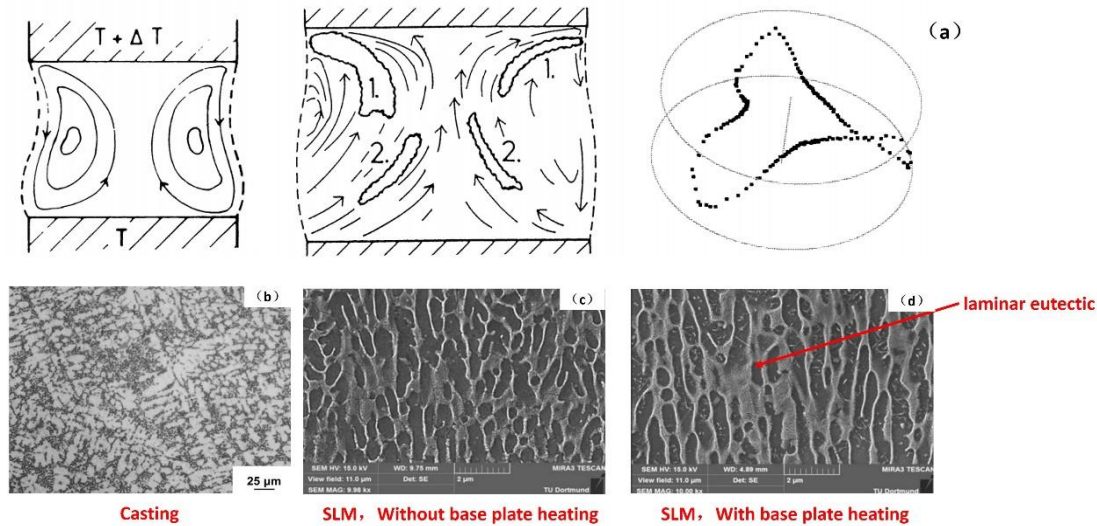

Fig. 20. The particle accumulation structures (PAS) in the ring vortex of thermocapillary flows are seen in (a) [70-72]. The microstructure by SEM of a casted Al-12Si alloy is shown in (b) [19], the microstructure after SLM of an Al-12Si alloy without preheating (c) and with preheating (d) [20].

## 5. Conclusions

(1) Single-layer and bulk 316L SLM experiments were presented in this paper to reveal the forming mechanism of sub-grain cellular microstructures in SLM. The submicron-scale cellular, elongated cellular or even band structures are always coexisting inside one single macro-solidified grain. Furthermore, the cellular structures are symmetrical with hexagonal, pentagonal and square cellular patterns where the cellular size is only around 1μm.

(2) The use of SLM creates ultra-high, nonlinear and asymmetrical temperature gradients and surface-tension gradients caused by non-uniform heating of free surface within the melt pool. Thermocapillary convection and intense convective vortex movements with high surface flow rates are all very important physical phenomena in SLM. These will form complicated flow patterns and influence the solidification microstructures observed.

(3) The unique sub-grain patterns are created by the complex temperature and surface

tension gradients and by the local convection and Bénard Instabilities in front of the solid/liquid (S/L) interface (so-called mushy zones). This nonlinear self-organization phenomenon (Bénard Instability) is superimposed on the macro-grain solidification to form the sub-grain patterns and microsegregations. More specifically, the solidification mode and morphologies of macro-grains are controlled by G and R, but the sub-grains patterns are controlled by flow instabilities and *Ma*. These two mechanisms are interacting and instability can always be generated in front of the moving S/L interface during SLM.

(4) The micro-segregations and distinct element distributions can be explained by the PAS mechanism; describing the behavior of small particles of a dilute concentration in a time-dependent (oscillatory) ring vortex thermocapillary flow. The austenitic solidification of 316L steel will reject the ferrite-promoting elements Cr, Si and Mo at the S/L interface and the enhanced capillary vortex will bring the rejected elements from to the instability layer. The two elements Si and Cr will react with residual oxygen and form the precipitated Cr-Si-O inclusions and the element Mo will enrich at the cellular boundaries by a Marangoni-Bénard convection and PAS mechanism. These explanations can also be used in Al-Si eutectic and CoCrMo prepared by SLM.

**Acknowledgement**

This work was supported by National Magnetic Confinement Fusion Science Program of China under Grant 2013GB109004 and 2014GB117000, and by National Natural Science Foundation of China under Grant 51361130032. The authors are grateful for the technical help and valuable discussions from Yuan Xue, Yi Zhai from Tsinghua University, and Dr. Thommy Ekström from Stockholm University.

# References:

[1] D.D. Gu, W. Meiners, K. Wissenbach, R. Poprawe, Laser additive manufacturing of metallic components: materials, processes and mechanisms, INT MATER REV 57 (2012) 133-164.

[2] X. Zhou, K. Li, D. Zhang, X. Liu, J. Ma, W. Liu, Z. Shen, Textures formed in a CoCrMo alloy by selective laser melting, J ALLOY COMPD 631 (2015) 153-164.

[3] J.P. Kruth, G. Levy, F. Klocke, T.H.C. Childs, Consolidation phenomena in laser and powder-bed based layered manufacturing, CIRP ANN-MANUF TECHN 56 (2007) 730-759.

[4] I. Yadroitsev, A. Gusarov, I. Yadroitsava, I. Smurov, Single track formation in selective laser melting of metal powders, J MATER PROCESS TECH 210 (2010) 1624-1631.

[5] L.E. Murr, E. Martinez, K.N. Amato, S.M. Gaytan, J. Hernandez, D.A. Ramirez, P.W. Shindo, F. Medina, R.B. Wicker, Fabrication of Metal and Alloy Components by Additive Manufacturing: Examples of 3D Materials Science, J. Mater. Res. Technol. 1 (2012) 42-54.

[6] B. SONG, X. ZHAO, S. LI, C. HAN, Q. WEI, S. WEN, J. LIU, Y. SHI, Differences in microstructure and proper ties between selective laser melting and traditional manufacturing for fabrication of metal par ts: A review, Front. Mech. Eng. 10 (2015) 111-125.

[7] L.E. Murr, S.M. Gaytan, D.A. Ramirez, E. Martinez, J. Hernandez, K.N. Amato, P.W. Shindo, F.R. Medina, R.B. Wicker, Metal Fabrication by Additive Manufacturing Using Laser and Electron Beam Melting Technologies, J MATER SCI TECHNOL 28 (2012) 1-14.

[8] L.E. Murr, Metallurgy of additive manufacturing: Examples from electron beam melting, Addit. Manuf. 5 (2015) 40-53.

[9] K. Saeidi, X. Gao, Y. Zhong, Z.J. Shen, Hardened austenite steel with columnar sub-grain structure formed by laser melting, MAT SCI ENG A-STRUCT 625 (2015) 221-229.

[10] I. Yadroitsev, P. Krakhmalev, I. Yadroitsava, S. Johansson, I. Smurov, Energy input effect on morphology and microstructure of selective laser melting single track from metallic powder, J MATER PROCESS TECH 213 (2013) 606-613.

[11] I. Yadroitsev, P. Krakhmalev, I. Yadroitsava, Hierarchical design principles of selective laser melting for high quality metallic objects, Addit. Manuf. 7 (2015) 45-56.

[12] M. Ma, Z. Wang, M. Gao, X. Zeng, Layer thickness dependence of performance in high-power selective laser melting of 1Cr18Ni9Ti stainless steel, J MATER PROCESS TECH 215 (2015) 142-150.

[13] G. Casalino, S.L. Campanelli, N. Contuzzi, A.D. Ludovico, Experimental investigation and statistical optimisation of the selective laser melting process of a maraging steel, OPT LASER TECHNOL 65 (2015) 151-158.

[14] K. Bartkowiak, S. Ullrich, T. Frick, M. Schmidt, New Developments of Laser Processing Aluminium Alloys via Additive Manufacturing Technique, LASERS IN MANUFACTURING 2011: PROCEEDINGS OF THE SIXTH INTERNATIONAL WLT CONFERENCE ON LASERS IN MANUFACTURING, VOL 12, PT A 12 (2011) 393-401.

[15] G.P. Dinda, A.K. Dasgupta, S. Bhattacharya, H. Natu, B. Dutta, J. Mazumder, Microstructural Characterization of Laser-Deposited Al 4047 Alloy, METALL MATER TRANS A 44A (2013) 2233-2242.

[16] K. Kempen, L. Thijs, J. Van Humbeeck, J.P. Kruth, Mechanical properties of AlSi10Mg produced by Selective Laser Melting, LASER ASSISTED NET SHAPE ENGINEERING 7 (LANE 2012) 39 (2012) 439-446.

[17] L. Thijs, K. Kempen, J. Kruth, J. Van Humbeeck, Fine-structured aluminium products with controllable texture by selective laser melting of pre-alloyed AlSi10Mg powder, ACTA MATER 61 (2013) 1809-1819.

[18] E. Brandl, U. Heckenberger, V. Holzinger, D. Buchbinder, Additive manufactured AlSi10Mg samples using Selective Laser Melting (SLM): Microstructure, high cycle fatigue, and fracture behavior,

MATER DESIGN 34 (2012) 159-169.

[19] K.G. Prashanth, S. Scudino, H.J. Klauss, K.B. Surreddi, L. Loeber, Z. Wang, A.K. Chaubey, U. Kuehn, J. Eckert, Microstructure and mechanical properties of Al-12Si produced by selective laser melting: Effect of heat treatment, MAT SCI ENG A-STRUCT 590 (2014) 153-160.

[20] S. Siddique, M. Imran, E. Wycisk, C. Emmelmann, F. Walther, Influence of process-induced microstructure and imperfections on mechanical properties of AlSi12 processed by selective laser melting, J MATER PROCESS TECH 221 (2015) 205-213.

[21] N. Read, W. Wang, K. Essa, M.M. Attallah, Selective laser melting of AlSi10Mg alloy: Process optimisation and mechanical properties development, MATER DESIGN 65 (2015) 417-424.

[22] L.N. Carter, C. Martin, P.J. Withers, M.M. Attallah, The influence of the laser scan strategy on grain structure and cracking behaviour in SLM powder-bed fabricated nickel superalloy, J ALLOY COMPD 615 (2014) 338-347.

[23] Y.S. Hedberg, B. Qian, Z. Shen, S. Virtanen, I.O. Wallinder, In vitro biocompatibility of CoCrMo dental alloys fabricated by selective laser melting, DENT MATER 30 (2014) 525-534.

[24] A. Takaichi, Suyalatu, T. Nakamoto, N. Joko, N. Nomura, Y. Tsutsumi, S. Migita, H. Doi, S. Kurosu, A. Chiba, N. Wakabayashi, Y. Igarashi, T. Hanawa, Microstructures and mechanical properties of Co–29Cr–6Mo alloy fabricated by selective laser melting process for dental applications, J MECH BEHAV BIOMED 21 (2013) 67-76.

[25] S. Bo, D. Shujuan, P. Coddet, L. Hanlin, C. Coddet, Fabrication of NiCr alloy parts by selective laser melting: Columnar microstructure and anisotropic mechanical behavior, MATER DESIGN 53 (2014) 1-7.

[26] Y. Lu, S. Wu, Y. Gan, J. Li, C. Zhao, D. Zhuo, J. Lin, Investigation on the microstructure, mechanical property and corrosion behavior of the selective laser melted CoCrW alloy for dental application, MAT SCI ENG C-MATER 49 (2015) 517-525.

[27] B. Vrancken, L. Thijs, J.P. Kruth, J. Van Humbeeck, Microstructure and mechanical properties of a novel beta titanium metallic composite by selective laser melting, ACTA MATER 68 (2014) 150-158.

[28] S. Kou, Welding Metallurgy, John Wiley & Sons, Incorporated, UNITED KINGDOM, 2003.

[29] I. Hemmati, V. Ocelik, J.T.M. De Hosson, Microstructural characterization of AISI 431 martensitic stainless steel laser-deposited coatings, J MATER SCI 46 (2011) 3405-3414.

[30] G.K. Lewis, E. Schlienger, Practical considerations and capabilities for laser assisted direct metal deposition, MATER DESIGN 21 (2000) 417-423.

[31] A.J. Pinkerton, L. Li, The effect of laser pulse width on multiple-layer 316L steel clad microstructure and surface finish, APPL SURF SCI 208 (2003) 411-416.

[32] J.D. Majumdar, A. Pinkerton, Z. Liu, I. Manna, L. Li, Microstructure characterisation and process optimization of laser assisted rapid fabrication of 316L stainless steel, APPL SURF SCI 247 (2005) 320-327.

[33] H. El Cheikh, B. Courant, S. Branchu, X. Huang, J. Hascoet, R. Guillen, Direct Laser Fabrication process with coaxial powder projection of 316L steel. Geometrical characteristics and microstructure characterization of wall structures, OPT LASER ENG 50 (2012) 1779-1784.

[34] M.A. Pinto, N. Cheung, M. Ierardi, A. Garcia, Microstructural and hardness investigation of an aluminum-copper alloy processed by laser surface melting, MATER CHARACT 50 (2003) 249-253.

[35] X. Lin, H. Yang, J. Chen, W. Huang, Mcriostructures evolution of 316L stainless steel during laser rapid forming, ACTA METALL SIN 42 (2006) 361-368.

[36] G. Phanikumar, P. Dutta, K. Chattopadhyay, Continuous welding of Cu-Ni dissimilar couple using

CO2 laser, SCI TECHNOL WELD JOI 10 (2005) 158-166.

[37] J.T. McKeown, A.K. Kulovits, C. Liu, K. Zweiacker, B.W. Reed, T. LaGrange, J.M.K. Wiezorek, G.H. Campbell, In situ transmission electron microscopy of crystal growth-mode transitions during rapid solidification of a hypoeutectic Al-Cu alloy, ACTA MATER 65 (2014) 56-68.

[38] Y. Guan, W. Zhou, H. Zheng, M. Hong, Y. Zhu, B. Qi, Effect of pulse duration on heat transfer and solidification development in laser-melt magnesium alloy, APPL PHYS A-MATER 119 (2015) 437-442.

[39] S.C. GILL, M. ZIMMERMANN, W. KURZ, LASER RESOLIDIFICATION OF THE AL-AL2CU EUTECTIC - THE COUPLED ZONE, ACTA METALLURGICA ET MATERIALIA 40 (1992) 2895-2906.

[40] R. Zhong, A. Kulovits, J.M.K. Wiezorek, J.P. Leonard, Four-zone solidification microstructure formed by laser melting of copper thin films, APPL SURF SCI 256 (2009) 105-111.

[41] W.J. BOETTINGER, D. SHECHTMAN, R.J. SCHAEFER, F.S. BIANCANIELLO, THE EFFECT OF RAPID SOLIDIFICATION VELOCITY ON THE MICROSTRUCTURE OF AG-CU ALLOYS, METALLURGICAL TRANSACTIONS A-PHYSICAL METALLURGY AND MATERIALS SCIENCE 15 (1984) 55-66.

[42] M. CARRARD, M. GREMAUD, M. ZIMMERMANN, W. KURZ, ABOUT THE BANDED STRUCTURE IN RAPIDLY SOLIDIFIED DENDRITIC AND EUTECTIC ALLOYS, ACTA METALLURGICA ET MATERIALIA 40 (1992) 983-996.

[43] A. KARMA, A. SARKISSIAN, INTERFACE DYNAMICS AND BANDING IN RAPID SOLIDIFICATION, PHYS REV E 47 (1993) 513-533.

[44] R.E. Rozas, A.L. Korzhenevskii, R. Bausch, R. Schmitz, Periodic layer formation in the growth of dilute binary alloys, PHYSICA A 413 (2014) 394-399.

[45] V.M. Azhazha, A.N. Ladygin, V.J. Sverdlov, P.D. Zhemanyuk, V.V. Klochikhin, Morphological transition in the cellular structure of single crystals of nickel-tungsten alloys near the congruent melting point, CRYSTALLOGR REP+ 501 (2005) S130-S135.

[46] D.G. BECK, S.M. COPLEY, M. BASS, THE MICROSTRUCTURE OF METASTABLE PHASES IN AG-CU ALLOYS GENERATED BY CONTINUOUS LASER MELT QUENCHING, METALLURGICAL TRANSACTIONS A-PHYSICAL METALLURGY AND MATERIALS SCIENCE 12 (1981) 1687-1692.

[47] M. Qu, L. Liu, Y. Cui, F. Liu, Interfacial morphology evolution in directionally solidified Al-1.5%Cu alloy, T NONFERR METAL SOC 25 (2015) 405-411.

[48] N. Chakraborty, The effects of turbulence on molten pool transport during melting and solidification processes in continuous conduction mode laser welding of copper-nickel dissimilar couple, APPL THERM ENG 29 (2009) 3618-3631.

[49] M. Asta, C. Beckermann, A. Karma, W. Kurz, R. Napolitano, M. Plapp, G. Purdy, M. Rappaz, R. Trivedi, Solidification microstructures and solid-state parallels: Recent developments, future directions, ACTA MATER 57 (2009) 941-971.

[50] E.L. Koschmieder, Bénard cells and Taylor vortices, Cambridge University Press, Cambridge [England], 1993.

[51] D. Schwabe, Convective instabilities in complex systems with partly free surface, Journal of Physics: Conference Series 64 (2007) 22.

[52] B. Kauerauf, G. Zimmermann, L. Murmann, S. Rex, Planar to cellular transition in the system succinonitrile-acetone during directional solidification of a bulk sample, J CRYST GROWTH 193 (1998) 701-711.

[53] G. Phanikumar, K. Chattopadhyay, P. Dutta, Modelling of transport phenomena in laser welding of dissimilar metals, INT J NUMER METHOD H 11 (2001) 156-174.

[54] J. Hu, H. Guo, H.L. Tsai, Weld pool dynamics and the formation of ripples in 3D gas metal arc welding, INT J HEAT MASS TRAN 51 (2008) 2537-2552.

[55] X. Zhou, D. Wang, X. Liu, D. Zhang, S. Qu, J. Ma, G. London, Z. Shen, W. Liu, 3D-imaging of selective laser melting defects in a Co–Cr–Mo alloy by synchrotron radiation micro-CT, ACTA MATER 98 (2015) 1-16.

[56] X. Zhou, Y. Tian, The simulation of 316L Selective Laser Melting: Thermoanalysis, Melt Flow and Residual Stresses., Tsinghua University, Beijing, 2015, p. 218.

[57] X. Bin, Z. Yuwen, Marangoni and buoyancy effects on direct metal laser sintering with a moving laser beam, Numerical Heat Transfer, Part A (Applications) 51 (2007) 715-733.

[58] Y. Zhao, Simulation of fluid dynamics behavior and solidified structure in welding pool, vol. Doctoral Thesis, Beijing University of Technology, Beijing, 2004.

[59] Z. Yuan, J. Ke, J. Li, The surface tension of metal and alloys, Science Press, Beijing, 2006.

[60] A.I. Mizev, D. Schwabe, Convective instabilities in liquid layers with free upper surface under the action of an inclined temperature gradient, PHYS FLUIDS 21 (2009) 112102.

[61] M. ASSENHEIMER, V. STEINBERG, TRANSITION BETWEEN SPIRAL AND TARGET STATES IN RAYLEIGH-BENARD CONVECTION, NATURE 367 (1994) 345-347.

[62] I. Ueno, T. Kurosawa, H. Kawamura, Thermocapillary Convection in Thin Liquid Layer with Temperature Gradient Inclined to Free Surface, 12th Int. Heat and Mass Transfer Conf., Grenoble, France, 2002.

[63] O.E. Shklyaev, A.A. Nepomnyashchy, Thermocapillary flows under an inclined temperature gradient, J FLUID MECH 504 (2004) 99-132.

[64] D.A. Egolf, I.V. Melnikov, W. Pesch, R.E. Ecke, Mechanisms of extensive spatiotemporal chaos in Rayleigh-Bernard convection, NATURE 404 (2000) 733-736.

[65] A. Mitchell, Fundamental development in electron-beam melting processes, Proceedings of the Electron Beam Melting and Refining State of the Art 1997 Conference, Reno, Nevada, USA, 1997.

[66] D. Schwabe, A.I. Mizev, M. Udhayasankar, S. Tanaka, Formation of dynamic particle accumulation structures in oscillatory thermocapillary flow in liquid bridges, PHYS FLUIDS 19 (2007).

[67] M.F. Schatz, S.J. VanHook, W.D. McCormick, J.B. Swift, H.L. Swinney, Time-independent square patterns in surface-tension-driven Benard convection, PHYS FLUIDS 11 (1999) 2577-2582.

[68] D.J. Kotecki, J.C. Lippold, Welding metallurgy and weldability of stainless steels, John Wiley, Hoboken, NJ, 2005.

[69] D. Kianersi, A. Mostafaei, J. Mohammadi, Effect of Welding Current and Time on the Microstructure, Mechanical Characterizations, and Fracture Studies of Resistance Spot Welding Joints of AISI 316L Austenitic Stainless Steel, METALL MATER TRANS A 45A (2014) 4423-4442.

[70] D. Schwabe, S. Tanaka, A. Mizev, H. Kawamura, Particle accumulation structures in time-dependent thermocapillary flow in a liquid bridge under microgravity, MICROGRAVITY SCI TEC 18 (2006) 117-127.

[71] D. Schwabe, S. Frank, Particle accumulation structures (PAS) in the toroidal thermocapillary vortex of a floating zone - Model for a step in planet-formation, in: A.C. LevasseurRegourd, J.C. Worms, D. Mohlmann, J. Klinger (Eds.), ADVANCES IN SPACE RESEARCH, vol. 23, 1999, pp. 1191-1196.

[72] D. Schwabe, Particle accumulation structures (PAS) in thermocapillary flow in floating zones, proceeding of 2nd European Symposium on the Utilisation of the International Space Station, ESTEC,

Noordwijk, The Netherlands., 1999.